\documentclass[conference,a4paper]{IEEEtran}
\IEEEoverridecommandlockouts

\usepackage{cite}
\usepackage{amsmath,amssymb,amsfonts}
\usepackage{algorithmic}
\usepackage{graphicx}
\usepackage{textcomp}
\usepackage{xcolor}
\usepackage{caption}
\usepackage{subcaption}
\usepackage{array}
\usepackage{wrapfig,lipsum,booktabs}
\usepackage[hidelinks]{hyperref}
\usepackage{fancyhdr}

\hypersetup{breaklinks=true}
\def\BibTeX{{\rm B\kern-.05em{\sc i\kern-.025em b}\kern-.08em
    T\kern-.1667em\lower.7ex\hbox{E}\kern-.125emX}}

\usepackage[font={small}]{caption}
\newenvironment{compactItemize}{\begin{list}{$\bullet$}
			{\setlength{\topsep}{1mm}\setlength{\itemsep}{0.25mm}
				\setlength{\parsep}{0.1mm}
				\setlength{\itemindent}{0mm}\setlength{\partopsep}{0mm}
				\setlength{\labelwidth}{15mm}
				\setlength{\leftmargin}{4mm}}}{\end{list}}
	
\newcounter{myctr}
\newenvironment{compactEnumerate}{\begin{list}{\arabic{myctr}.}
        {\usecounter{myctr}
            \setlength{\topsep}{1mm}\setlength{\itemsep}{0.5mm}
            \setlength{\parsep}{0.0mm}
            \setlength{\itemindent}{0mm}\setlength{\partopsep}{0mm}
            \setlength{\labelwidth}{15mm}
            \setlength{\leftmargin}{4mm}}}{\end{list}}

\begin{document}

\fancypagestyle{firststyle}
{
   \fancyhf{}
   \fancyhead[C]{This is an authors' copy of the paper  to appear in the 2026 Design, Automation \& Test in Europe Conference (DATE).}
   \renewcommand{\headrulewidth}{0pt} 
}

\title{Cyber-Physical System Design Space Exploration for Affordable Precision Agriculture\vspace{-10pt}}

\author{\IEEEauthorblockN{Pawan Kumar}
\IEEEauthorblockA{\textit{Arizona State University} \\
pkumar97@asu.edu}
\and
\IEEEauthorblockN{Hokeun Kim}
\IEEEauthorblockA{\textit{Arizona State University} \\
hokeun@asu.edu}
}

\IEEEaftertitletext{\vspace{-2.0\baselineskip}}
\maketitle

\thispagestyle{firststyle}

\begin{abstract}
Precision agriculture promises higher yields and sustainability, but adoption is slowed by the high cost of cyber-physical systems (CPS) and the lack of systematic design methods. 
We present a cost-aware design space exploration (DSE) framework for multimodal drone–rover platforms to integrate budget, energy, sensing, payload, computation, and communication constraints. 
Using integer linear programming (ILP) with SAT-based verification, our approach trades off among cost, coverage, and payload while ensuring constraint compliance and a multitude of alternatives.
We conduct case studies on smaller and larger-sized farms to show that our method consistently achieves full coverage within budget while maximizing payload efficiency, outperforming state-of-the-art CPS DSE approaches.
\end{abstract}

\section{Introduction}

Modern agriculture faces challenges to sustainably feed a growing global population amid workforce shortages, climate variability, and constrained natural resources~\cite{McFadden2023}. 
Precision agriculture has emerged to tackle these issues with advanced sensing, automation, and data-driven management to optimize crop production. 
In particular, cyber-physical systems (CPS) integrate with sensors, microcontrollers, and real-time image processing, enabling continuous, non-destructive monitoring of crop growth and targeted interventions to boost yields while keeping hardware costs moderate~\cite{Kumar2024,Castillejo2020}.
Such CPS-driven precision agriculture not only improves resource use efficiency (e.g., minimizing usage of water, fertilizer, and pesticide) but also helps offset labor constraints and environmental impacts, allowing farmers to achieve higher yields more sustainably~\cite{Pretto2021}. 

\begin{figure}
\centering
	\includegraphics[width=0.65\columnwidth]{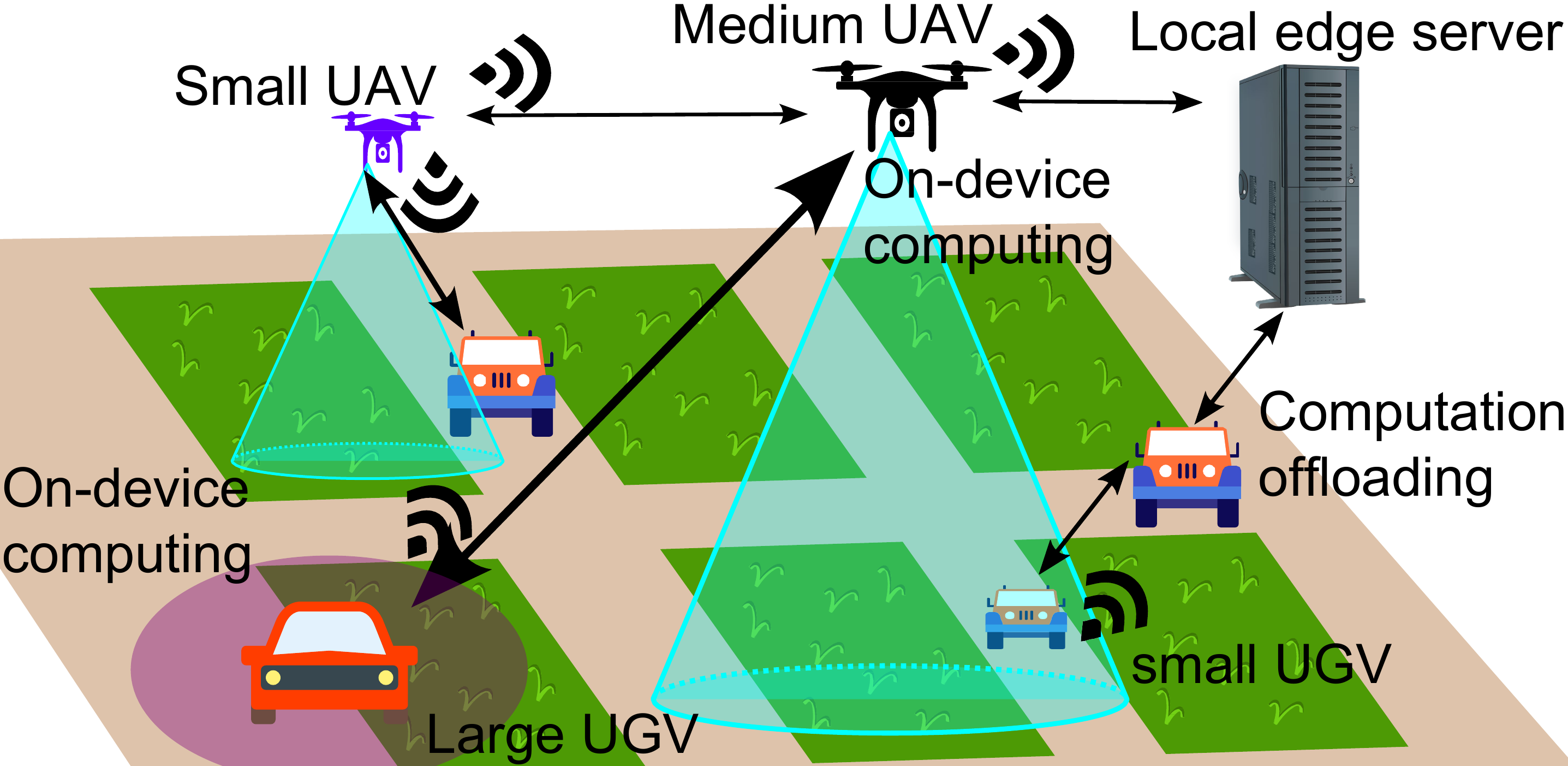}
\vspace{-5pt}
	\caption{ Example illustration of DSE for cost-effective multimodal software-hardware platforms with unmanned ground vehicles (UGVs or rovers) and unmanned aerial vehicles (UAVs or drones) for precision agriculture.}
    \vspace{-20pt}
	\label{fig:Intro}
\end{figure}

Design space exploration (DSE) is critical for developing cost-effective CPS in precision agriculture, but it is challenging due to conflicting goals of minimizing cost while ensuring adequate coverage, payload, and runtime~\cite{Madroal2021,Pretto2021}.
Balancing these trade-offs requires systematic methods to explore hardware–software configurations under budget and operational constraints. 
Recent work emphasizes multimodal platforms that maximize field coverage and enable targeted interventions for sustainability and reduced chemical usage~\cite{Kumar2025,USGAO2024}.
This synergy between DSE and multimodal platforms provides a scalable pathway to practical, data-driven precision agriculture. \figureautorefname~\ref{fig:Intro} shows an integrated multimodal platform hardware configuration, 
illustrating the trade-offs among size, coverage, and computing capacity to be evaluated.

\subsection{Related Work and State-of-the-Art Approaches}
Recent studies have introduced cost-effective CPS prototypes and predictive modeling to non-destructively estimate plant growth, thereby promoting sustainability and improving crop yields~\cite{Bouarourou2024,Kumar2024,Qiu2024,Shilpa2024}. 
Doering \textit{et al.}~\cite{Doering2014} propose a UAV-based platform with adaptive mission planning for precision agriculture. 
These methods integrate CPS in precision agriculture to facilitate data acquisition and adaptive decision-making in response to dynamic environmental conditions.

DSE is crucial in optimizing CPS configurations by balancing performance, cost, and computational efficiency. 
M\"{u}hleis \textit{et al.}~\cite{muhleis2011co} integrate control performance directly into DSE.
Fitzgerald \textit{et al.}~\cite{Fitzgerald2015} develop a toolchain unifying discrete and continuous CPS models. 
Herget \textit{et al.}~\cite{Herget2022} improve workload modeling for distributed CPS.  
Liao~\textit{et~al.}~\cite{Liao2023} refine subspace pruning to cut computational overhead. 
Xiao~\textit{et~al.}~\cite{xiao2024} propose ContrArc, which uses contract-based pruning to eliminate infeasible architectures before exploration.
%
Yu \textit{et al.}~\cite{Yu2023} explore symbiotic CPS designs using human-in-the-loop feedback to iteratively refine UAV architectures.
%
Zheng \textit{et al.}~\cite{Zheng2022} develop a gradient-free framework for cross-domain UAV design exploration.
There is, however, a critical gap that none of these methods target CPS DSE for precision agriculture.
This motivates the need for a domain-aware CPS DSE framework tailored to the constraints and goals of agricultural automation.

\subsection{Research Problem and Contributions}

A unique challenge in the agricultural CPS  DSE problem is the co-parameterization of crops and applications, with non-linear relationships between crops and sensing/coverage requirements.
Specifically, developing CPS for agriculture requires balancing competing objectives: minimizing cost while ensuring sufficient sensing, computation, and field coverage. 
Current state-of-the-art methods are limited to optimizing individual factors, leaving farmers without a systematic way to evaluate full platform designs within budget and constraints.

We address this gap with a cost-aware DSE framework that evaluates heterogeneous multimodal platform configurations across budget, energy, and sensing constraints. 
The underlying integer linear programming (ILP) formulation is non-trivial because many of these relationships are inherently non-linear, requiring careful modeling and relaxation to obtain feasible solutions. 
To guarantee feasible outputs, we complement our approach with SAT-based verification to check the generated platforms and filter out invalid ones.
Our framework produces alternative designs that achieve high coverage and sensing performance at the lowest possible cost, providing a principled method to navigate complex design trade-offs. 
We release the design, implementation, data sets, and evaluation results of our approach through an open-source repository.\footnote{\href{https://github.com/asu-kim/cps-dse-apa}{https://github.com/asu-kim/cps-dse-apa}}

\section{Proposed Approach}
This section outlines our proposed DSE method
with explanations of user inputs and hardware factors to be explored, leading to feasible platforms for agriculture.
\subsection{User Inputs and Design Considerations} \label{sec:user_inputs}
Users first provide key inputs that shape the design space: the total available budget, the size of the farm area, the target crop type, and the desired applications, as shown below.

\begin{compactEnumerate}
\item \textbf{Budget:} Sets an upper limit on the total allowable monetary cost for hardware configurations, ensuring that our platforms are practical for farms with varying financial constraints.

\item \textbf{Farm area size:} Determines the minimum area coverage requirement that the platform must achieve. 
Larger farms may demand multiple vehicles or higher-capacity components to maintain efficiency.

\item \textbf{Target crop:} 
Determines the suitability of drones, rovers, or both for field operations. Indoor crops grown in controlled environments typically favor rovers, as drones are impractical in confined spaces. Densely planted crops like fiber and legumes increase the risk of ground vehicle damage, making drones a better choice. Similarly, early-stage paddy fields are often waterlogged, limiting rover use and favoring aerial monitoring. In contrast, crops such as cereals, trees, orchards, vines, vegetables, forage, and oilseeds benefit from hybrid drone–rover platforms, combining aerial surveys with ground-level intervention.
\item \textbf{Application:} Defines exact tasks that the platform must perform. 
\tablename~\ref{table:applications} lists various agricultural applications that users can choose from when setting up our DSE framework. 
Each application has a recommended processing mode, either off-board (using a local edge server) or onboard (computing directly on the vehicle), and specifies whether it is better suited to drones, rovers, or both.
Applications requiring control and actuation are handled onboard, while others are off-board.
\end{compactEnumerate}

\noindent
Once these inputs are defined, our DSE framework systematically translates them into the design considerations that determine which combinations of chassis, motors, batteries, computing units, and sensors can meet the farm’s needs within the budget and platform constraints. It then evaluates each candidate using five key metrics:
\begin{compactEnumerate}
    
    \item \textbf{Cost:} Sum of base vehicle, component (e.g., LiDAR, manipulators), and edge-server costs.
    \item \textbf{Coverage:} Maximum operational coverage area (m\textsuperscript{2}), calculated by estimating available battery energy after accounting for peak draws from motors, sensors, and onboard processors, then integrating over average platform speed.
    \item \textbf{Payload capacity:} Mass (kg) available for additional components such as cameras, sensors, or manipulators, varying according to the target crop and application.
    \item \textbf{Runtime:} Time spent (hours) before recharge, computed under the same maximum-component-load assumption.
    \item \textbf{Crop–Application Parameterization:} Parameters tailored to specific crop types and application scenarios. 
\end{compactEnumerate}

\begin{table}
\vspace{5pt}
\scriptsize
\centering
\begin{tabular}{|p{3.6cm}|p{2.1cm}|p{0.8cm}|}
\hline
\textbf{Application} & \textbf{Computation Mode} & \textbf{Platform} \\
\hline
General crop monitoring & Off-board & Both \\
Thermal imaging & Off-board & Both \\
Image stitching & Off-board & Both \\
Soil monitoring & Off-board & Rover \\
Yield estimation & Off-board & Both \\
Quality control & Off-board & Rover \\
Autonomous fruit/vegetable picking & Onboard & Rover \\
Mechanical weeding & Onboard & Rover \\
Soil pH sampling & Onboard & Rover \\
Climate mapping & Off-board & Rover \\
Fence/infrastructure inspection & Off-board & Drone \\
Livestock monitoring & Onboard & Both \\
Beehive inspection & Onboard & Rover \\
Frost \& pest early-warning systems & Onboard & Both \\
Fertilizing & Onboard & Rover \\
\hline
\end{tabular}
\caption{Overview of precision agriculture applications and their recommended processing modes and hardware platform compatibility with rover, drone, or both.}
    \vspace{-20pt}
	\label{table:applications}
\end{table}

Our DSE framework encodes budget, area coverage, runtime thresholds, payload, communication-range requirements, and crop–application specifications into ILP.
The goal of the objective function in Eq.\ref{eq:Objective_function} is to minimize a weighted combination of normalized cost, area coverage, and payload.
By varying priority weights (\(\alpha, \beta, \gamma\)), our ILP solver systematically generates all the possible drone-rover configurations that pass all constraints. 
 Then, it screens them for feasibility using the crop–application platform compatibility and component power limits,
\vspace{-5pt}
\begin{equation}
\min_{x}\;\sum_{i=1}^{N} \Bigl(\alpha\,\tilde C_i\,x_i + \beta\,\tilde A_i\,x_i - \gamma\,\tilde P_i\,x_i\Bigr)
\label{eq:Objective_function}
\vspace{-6pt}
\end{equation}
$N$ denotes the total number of possible configurations, and $i$ is the index of each configuration. 
For configuration $i$, $\tilde{C}_i$ is the normalized cost, $\tilde{A}_i$ is the area coverage, and $\tilde{P}_i$ is the per-platform payload.
The variable $x_i$ indicates the number of units selected for configuration $i$. 
Finally, $\alpha$, $\beta$, and $\gamma$ are weight factors used to balance cost, area coverage, and payload capacity using the 
ROC weighting scheme described in ~\ref{Weighted Score Design}.




\subsection{Design Space Exploration}
In this section, we formalize the framework requirements through a series of linear inequalities that encode budget limits, coverage and payload needs, runtime targets, component sizing rules, and communication-range expectations. 
These constraints approximate relationships that behave non-linearly in practice, such as the way weight influences energy consumption, how motor torque scales with usable payload, and how coverage depends on battery discharge and geometric range.
Together, they define the feasible design space that our ILP solver searches to find valid drone-rover platforms.

We use equation (Eq.~\ref{eq:Budget_constraint}) to ensure the total cost does not exceed the user’s budget,
\begin{equation}
(C_i + extra\_cost) \cdot x_i + edge\_cost \leq B
\label{eq:Budget_constraint}
\vspace{-6pt}
\end{equation}
Here, $C_i$ is the base monetary cost of configuration $i$, which includes the vehicle chassis, core hardware, and essential built-in components. 
$B$ denotes the total user-defined budget. 
The term $extra\_cost$ accounts for additional monetary cost for application-specific parts, while $edge\_cost$ represents the monetary cost of the local edge server.

Eq.~\ref{eq:Area_constraint} ensures selected rovers and drones cover at least the required farm size,
\begin{equation}
A_i \cdot x_i \geq S
\label{eq:Area_constraint}
\vspace{-6pt}
\end{equation}
Here, $S$ denotes the farm size specified by the user, $x_i$ is the number of units selected for configuration $i$, and $A_i$ represents the coverage area each unit of configuration $i$ can achieve.

To determine how much area each selected configuration can cover, we use
Eq.\ref{eq:Area_calculate} to find the operational coverage area for any drone or rover configuration in our DSE,
\begin{equation}
A_i = BATT_{i} \cdot T_{i} \cdot  W_{\text{component},i} \cdot f_{\text{weight},i}(W_{\text{max}, i})
\label{eq:Area_calculate}
\end{equation}
Here,
$BATT_i$ and $T_i$ denote the number of batteries and tires in configuration $i$. 
$W_{\text{component},i}$ is the weight calculated in Eq.~\ref{eq:carry_weight_rover}
$W_{\text{max},i}$ is the maximum payload of configuration $i$ calculated in Eq.~\ref{eq:Max_weight_rover}, and $f_{\text{weight},i}(W_{\text{max},i})$ captures the efficiency reduction due to weight, as defined in Eq.~\ref{eq:Weight_area_factor}.
\begin{equation}
f_{\text{weight},i}(W_{\text{max},i}) = \frac{1}{1 + W_{\text{max},i}/100}
\label{eq:Weight_area_factor}
\end{equation}

We use Eq.~\ref{eq:Force_motors} to find the total force the motors generate, which determines how much weight the platforms can move,
\begin{equation}
W_{\text{carry}} = \tau \times \frac {nm} {TR \cdot g}
\label{eq:Force_motors}
\vspace{-6pt}
\end{equation}
Here, $W_{\text{carry}}$ is the effective motor weight (in kg), computed by dividing the total torque-based force $\tau$ by the tire radius $TR$, and the gravitational acceleration $g$. 
The term $nm$ denotes the number of motors in configuration $i$. 
For drone configurations, tire radius is omitted from the computation of $W_{\text{motors}}$.

We use Eq.~\ref{eq:Max_weight_rover} and Eq.~\ref{eq:carry_weight_rover} to find the maximum payload of configuration $i$.
\begin{equation}
W_{{max},i} = W_{\text{carry},i} - W_{\text{component},i}
\label{eq:Max_weight_rover}
\end{equation}
\begin{equation}
W_{\text{component},i} = W_{\text{chassis},i} + W_{\text{motor},i} + W_{\text{battery},i} + W_{\text{tires},i}
\label{eq:carry_weight_rover}
\end{equation}

Here, 
$W_{\text{component},i}$ is the sum of physical weights.
$W_{\text{chassis},i}$, $W_{\text{battery},i}$, $W_{\text{motors},i}$, and $W_{\text{tires},i}$ represent the physical weights of the chassis, battery, motors, and tires, respectively of configuration $i$. 
For drone configurations, tire weight is omitted from the computation of $W_{\text{component}}$.


We use the following equation, Eq.~\ref{eq:Battery_rover}, to ensure each selected drone or rover configuration includes at least one battery module to meet its energy requirements.
\begin{equation}
\sum_s BATT_{i,s} \geq 1, \quad \forall i
\label{eq:Battery_rover}
\end{equation}
Here, $BATT_{i,s}$ is a binary variable indicating whether battery option $s$ is included in configuration $i$, and $s$ represents the available battery choices for rover configurations.

Eq.~\ref{eq:Rover_size_match} ensures each motor matches the chassis size, maintaining power-to-weight balance for reliable operation.
\begin{equation}
M_{\text{size},i}\;=\;CH_{\text{size},i}, \quad \forall i
\label{eq:Rover_size_match}
  \vspace{-6pt}
\end{equation}
Here, $M_{\text{size},i}$ and $CH_{\text{size},i}$ denote the motor size and chassis size, respectively, for configuration $i$.

We observe that drone configurations typically have a much shorter flight time than rovers due to higher power draw and limited battery capacity. To balance this gap and ensure that drones can perform meaningful tasks without frequent recharging, we add Eq.~\ref{eq:Runtime} below as a flight time condition.
\begin{equation}
\quad f_d \ge F_{\min}
\label{eq:Runtime}
  \vspace{-6pt}
\end{equation}
Here, $f_d$ denotes the estimated runtime (flight time in hours) of a specific drone configuration $d$, and $F_{\min}$ is the minimum required flight time per unit, set to 0.2 hours. This threshold reflects the flight time of commercial drones such as the Flyability Elios 3~\cite{FlyabilityElios3} and XAG V40~\cite{XAGV40}, which typically operate for about 0.2 hours.

We ensure the drone-rover configurations maintain full communication coverage across the farm.
Eq.~\ref{eq:Comm_range} below calculates the number of cells\footnote{We use the term ``cell'' to refer to the area covered by each device of the communication method, analogous to cells in cellular networks.} by dividing the farm area by the communication method’s effective range.
\begin{equation}
\begin{aligned}
l =\frac{\sqrt{S}}{R_{\mathrm{comm}} \times 1000}
 \end{aligned}
  \label{eq:Comm_range}
  \vspace{-6pt}
 \end{equation}
Here, $S$ is the total farm size specified by the user in m\textsuperscript{2} (square meter), $R_{\text{comm}}$ is the range of a communication method (e.g., LoRa, cellular, Wi-Fi) in km, and $l$ is the number of cells (minimum of 1 cell) required to cover the entire area.
 Eq.~\ref{eq:Comm_range_2} then multiplies the number of cells from  Eq.~\ref{eq:Comm_range} by the cost per cell to compute the total communication cost.
 \begin{equation}
 C_{\text{comm}} = l\times C_{\text{cell}}
 \label{eq:Comm_range_2}
  \vspace{-6pt}
 \end{equation}
Here, $C_{\text{cell}}$ denotes the cost of a single communication cell, and $C_{\text{comm}}$ represents the total communication cost.

A suitable computing unit is selected based on the application's processing requirements as shown in (\tablename~\ref{table:applications}).
On-board processing requires real‑time tasks such as image processing or autonomous navigation, so we assign a GPU or TPU. For off-board processing, the configuration may instead use a CPU, GPU, or TPU.
Computation performance is evaluated using Geekbench 5.4.1~\cite{Geekbench5,Geekbench541,chetoui2021} multi-core scores, which are derived from workload execution time, the lower the execution time higher the score. 


\subsection{SAT-based Verification} \label{SAT based Verification}

To ensure the correctness of the ILP even after modification of the ILP formula by the users,
we verify that each configuration meets the total cost and total coverage constraints using a SAT solver, PySAT~\cite{Py-sat}.
We encode feasibility as Boolean variables, enabling efficient SAT solving. 
The solver checks conditions in Eq.~\ref{eq:SAT}.
 \vspace{-6pt}
\begin{equation}
\text{total\_cost} \leq B, \quad \text{total\_coverage} \geq S
 \label{eq:SAT}
 \vspace{-3pt}
\end{equation}
where $B$ is the budget and $S$ the required farm size.
Hardware components (motors, batteries, sensors, frames) contribute to these totals, keeping verification aligned with design.
Similar SAT-based methods are used in many-core DSE~\cite{richthammer2018},~\cite{schwarzer2017}, validating our approach for CPS.
\section{Evaluation}
To assess the effectiveness of our approach, we first evaluate the design aspects considered in our approach. Second, we compare our approach's optimization method against state-of-the-art approaches.
\begin{compactItemize}
    \item \textbf{Evaluation of design aspects (Section~\ref{sec:eval_design_aspects}) --} We conduct an ablation study of the ILP-based multi-objective optimizers: \textit{Proposed Approach (\texttt{PA}) (cost+area+payload)}, \textit{Area+Payload} \texttt{AP}, \textit{Cost+Area} \texttt{CA}, and \textit{Payload+Cost} (\texttt{PC}).
    \item \textbf{Optimizer model evaluation (Section~\ref{sec:opt_eval}) --} Comparison of the proposed ILP approach against optimization methods existing state-of-the-art CPS DSE approaches including: \textit{simulated annealing}, \textit{Bayesian optimization (BO)}~\cite{Yu2023}, \textit{random search}, \textit{genetic algorithm}, \textit{portfolio optimization}, \textit{discrete search}, \textit{Lengler}~\cite{Zheng2022}, and \textit{PG-DSE with PSP and PEGA}~\cite{Liao2023} as summarized in \tablename~\ref{table:evaluation_table}.
\end{compactItemize}
Our evaluation uses two farm settings as case studies:

\noindent
\textbf{Case Study 1}:
We assume a smaller-scale farm with a farm size of 4047 m$\textsuperscript{2}$ (approx. 1 acre), which aligns with farm size distribution data published by the USDA NASS in the 2022 Census of Agriculture~\cite{USDAfarm} and by the EU in 2020~\cite{EUfarm}, with a budget of \$100K (USD), which aligns with entry-level costs for robotic platforms used in precision farming \cite{AllyNav, Gil2023}.
We select the tree crop (Section~\ref{sec:user_inputs}) for its high monitoring demands and economic value, as shown in reports from the US and Europe, with the value of \$2.9 billion USD in 2024~\cite{USDAcrop} for apple, and 1.17 million acres of farm land used for growing apples~\cite{EUapple}.
We use autonomous fruit and vegetable picking (\tablename~\ref{table:applications}) as applications leveraging CPU, GPUs, or TPUs for efficient harvesting tasks on rovers.

\noindent
\textbf{Case Study 2}:
We assume a large-scale farm with a farm size of 40,469 m$\textsuperscript{2}$ (approx. 10 acres)~\cite{USDAfarm,EUfarm}, with a budget of \$1M (USD), which aligns with the cost for robotic weeding machines \cite{AllyNav, Bloch2019, Gil2023}.
We select vine (Section~\ref{sec:user_inputs}) due to its high value and requirement for delicate handling, as shown in reports from the US and Europe, with the value of \$6.2 billion USD in 2024~\cite{USDAcrop} for grapes, and 7.9 million acres of farm land used for growing grapes~\cite{EUapple}.
We use general crop monitoring and yield estimation  (\tablename~\ref{table:applications}), which uses off-board processing with CPU, GPU, or TPU.

Across both evaluations, we visualize scatter plots and compute a unified weighted score for each optimizer. 
\subsection{Evaluation of Design Aspects}
\label{sec:eval_design_aspects}
\begin{figure}
\centering	\includegraphics[width=1.0\columnwidth]{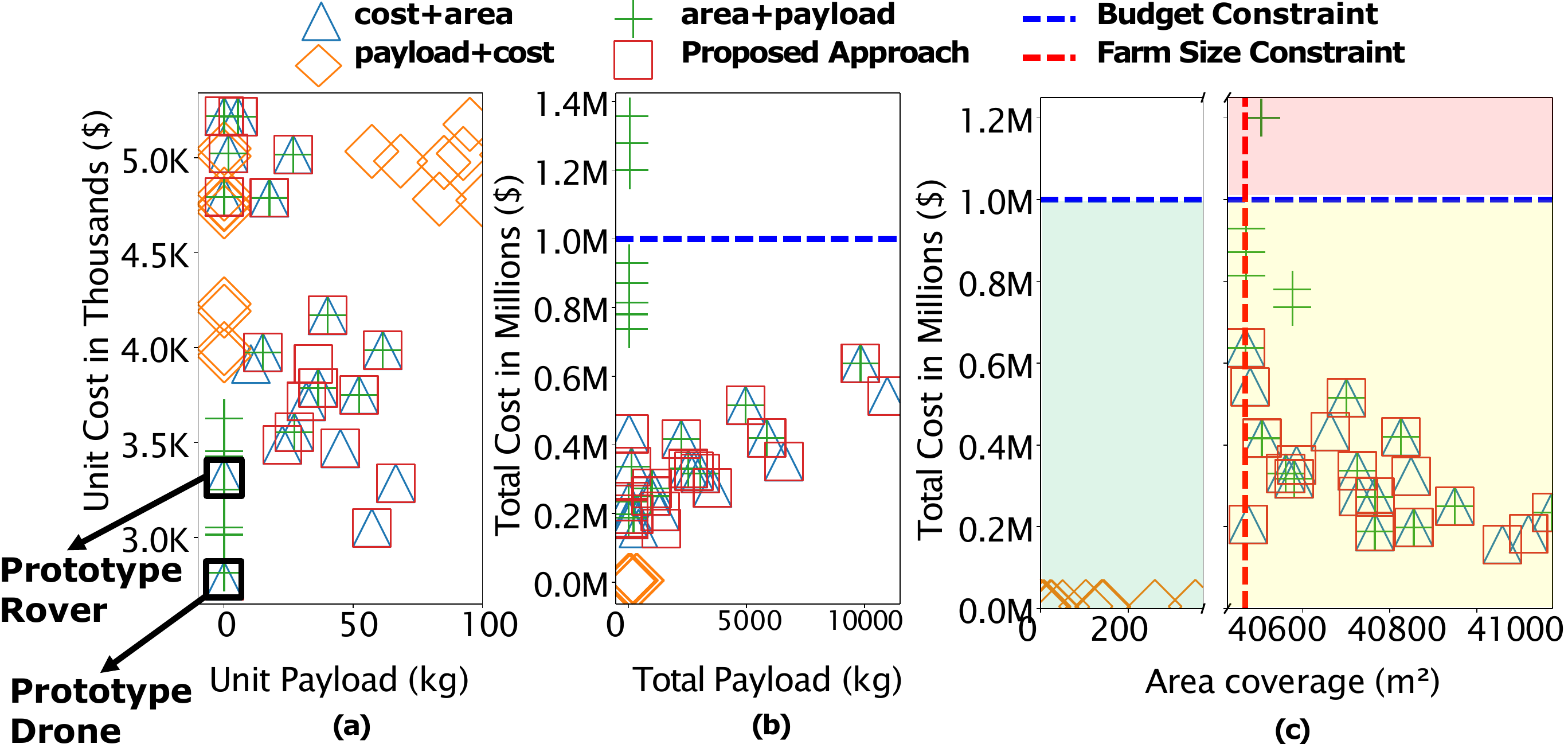}
\vspace{-5pt}
	\caption{Trade-offs among design aspects in Case Study 2 (\$1M, 10 acres): (a) unit cost vs. unit payload (b) total cost vs. total payload, (c) total cost vs. area coverage.}
    \vspace{-10pt}
	\label{fig:self_frontier_B1}
\end{figure}


Figure~\ref{fig:self_frontier_B1} illustrates trade-offs across cost, payload, and area coverage for different optimization methods for Case~Study~2. 
A unit refers to a single drone or rover instance, and a configuration may include multiple units to satisfy minimum farm area coverage.
\figurename~\ref{fig:self_frontier_B1}--(a) plots the unit cost vs. unit payload where,  \texttt{PA} and \texttt{CA} yield balanced designs, whereas \texttt{PC} increases cost with increase in payload. 
\texttt{AP} produces a more scattered distribution. 
\figurename~\ref{fig:self_frontier_B1}--(b) plots the total cost vs. total payload where, \texttt{PC} clusters near the origin, while \texttt{PA} and \texttt{CA} achieve high payloads within \$1M. Whereas, \texttt{AP} has lower payload and also exceeds the \$1M.
\mbox{\figurename~\ref{fig:self_frontier_B1}--(c)} plots the total cost vs. area coverage where, \texttt{PA} and \texttt{CA} maximize area coverage without exceeding \$1M (yellow shaded region). 
In contrast, \texttt{AP} attains minimum coverage but includes configurations that exceed \$1M (red shaded region), whereas \texttt{PC} stays under \$1M but fails coverage (green shaded region).

\noindent
\textbf{Weighted Score Design}:
\label{Weighted Score Design}
We prioritize cost to achieve our primary goal, a cost-effective DSE. 
We assign area coverage the next highest priority so our framework can monitor the minimum required area for the applications. 
We consider payload as the least critical factor once the drone–rover carries the necessary sensors, as extra capacity adds little value.
To reflect this, we now use the Rank Order Centroid (ROC)~\cite{HuttonBarron1992} method to determine the final weights, following the approach demonstrated in the works~\cite{Varshney2024, Roszkowska2013, Suhendarwan2024}.
Using ROC, we rank the criteria in order of importance and derive their normalized weights as follows:
\begin{wraptable}{r}{5.8cm}
\vspace{-10pt}
    \footnotesize
	\centering
	\begin{tabular}{|p{0.922cm}|>{\centering\arraybackslash}p{1.75cm}|>{\centering\arraybackslash}p{1.8cm}|}
		\hline
		\textbf{Design ~~~Aspects} & \textbf{Case Study 1 \$100K, 1 Acre} & \textbf{Case Study 2 \$1M, 10 Acres} \\
		\hline
		\texttt{AP} & 0.404 & 0.377 \\
        \hline
        \texttt{PC} & 0.611 & 0.611  \\
        \hline
		\texttt{CA} & 0.417 & 0.389  \\
        \hline
		\texttt{PA} & 0.417 & 0.389 \\
		\hline
	\end{tabular}
    \vspace{-5pt}
    	\caption{Weighted-score comparison for design aspects in our case.}
    \vspace{-15pt}
	\label{table:self_W}
\end{wraptable}

\begin{table*}
\vspace{5pt}
	\centering
    \scriptsize
	\begin{tabular}{|p{1.2cm}|p{1.3cm}|p{1cm}|p{4.3cm}|p{3.2cm}|p{2.1cm}|p{2cm}|}
		\hline
		\textbf{Approach} & \textbf{Domain} & \textbf{Platform} & \textbf{Evaluation Methodology} & \textbf{DSE Formulation} & \textbf{Optimizers} & \textbf{Cost Breakdown} \\
		\hline
		DESTION 23~\cite{Yu2023}  & General CPS & UAVs \& Robot Car & Simulation-based performance evaluations include time \& energy for robot car \& four flight tasks for UAVs & Reward feedback \& switching strategy & Simulated Annealing, Bayesian Optimization (BO) & Not ~~available \\
        \hline
		DESTION 22~\cite{Zheng2022} & UAV architecture & UAVs only & 4 mission-based benchmarks (hover, straight, circle, oval) & Graph-based design  optimized by component selection or action sequences & Discrete, Lengler, Portfolio, FastGA, Random Search & Not ~~available \\
        \hline
		ASP-DAC 23~\cite{Liao2023}  & ADAS & FPGA-Xilinx Artix-7 & ADAS system: 10 PSMs, 52 MCCs, 244 MCC alternatives, \& Synthetic systems with design space sizes & PSMs leverage component-level MCCs to model time-triggered behavior & PG-DSE with PSP and PEGA & Not ~~available \\
        \hline
        Proposed approach & Precision agriculture & Rovers \& UAVs & Evaluation of design aspects \& Optimizer model evaluation & Minimize cost, maximize area coverage \& payload & Integer Linear Programming & Available \\
		\hline
	\end{tabular}
	\caption{Comparison of the proposed and state-of-the-art CPS DSE approaches.}
    \vspace{-15pt}
	\label{table:evaluation_table}
\end{table*}

\begin{compactItemize}
\vspace{-15pt}
    \item Weight for cost: 0.611
    \item Weight for area coverage: 0.278
    \item Weight for payload: 0.111
\end{compactItemize} 
Under these weights for Case Study 2, \tablename~\ref{table:self_W} shows that the \texttt{PC}  achieves the highest score of 0.611, followed by \texttt{PA} and \texttt{CA} at 0.389.
\texttt{AP} ranks lowest at 0.377.

A similar pattern appears in Case Study 1. 
\texttt{PA} and \texttt{CA} again achieve full area coverage within \$100K, while \texttt{PC} stays with the highest weighted score but lacks full area coverage, and \texttt{AP} exceeds \$100K despite minimum area coverage. 
Together, we confirm that the \texttt{PA} and \texttt{CA} deliver the best balance of cost and area coverage across different farm sizes.

To summarize,
\figurename~\ref{fig:self_frontier_B1} and \tablename~\ref{table:self_W} confirm that the \texttt{PA} and \texttt{CA} achieve full area coverage while keeping costs within the \$1M budget in Case Study 2, which explains why they earn balanced weighted scores. 
Meanwhile, the \texttt{PC} scores higher due to its low cost and high payload, but it fails to fully meet the area coverage requirement. 
\texttt{AP} exceeds the budget even though it meets the area coverage target and ranks lowest in the weighted score. 
\subsection{Optimizer Model Evaluation}
\label{sec:opt_eval}

\begin{figure}
\centering
\includegraphics[width=0.9\columnwidth]{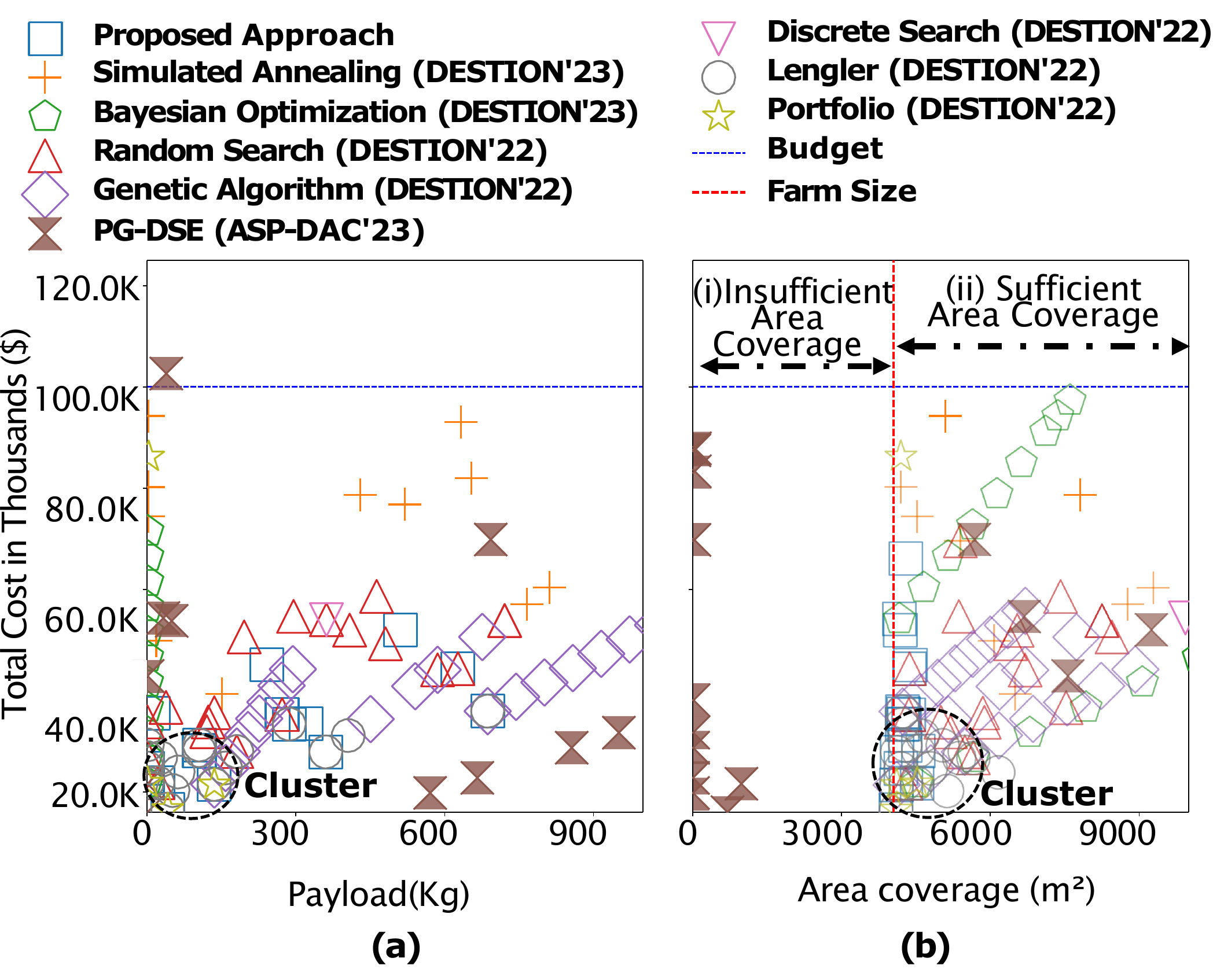}
\vspace{-5pt}
	\caption{Comparative evaluation of optimizers in Case Study 1 (\$100K, 1 acre): (a) total cost vs. payload, (b) total cost vs. area coverage.}
    \vspace{-10pt}
	\label{fig:model_frontier_B2}
\end{figure}

\figurename~\ref{fig:model_frontier_B2}--(a) plots the total cost vs. payload for Case Study 1, where the \$100K budget line appears as a blue dotted line.
The \texttt{PA} spans a broad payload range while remaining below \$100K, indicating flexible scaling across configurations.
\textbf{simulated annealing} concentrates around mid-to-high (approx. 400-800 kg) payload values within \$100K.
\textbf{genetic algorithm} and \textbf{random search} show a gradual cost increase as payload increases.
\textbf{discrete search} shows only one configuration at 400 kg payload with a cost of 60K.
\textbf{BO} shows an increase in costs without huge changes in payload.
\textbf{PG-DSE} covers a wide span, from light payload (approx. 50 kg) at mid cost (approx. \$60K) to heavy payload (approx. 900 kg) at low cost (approx. \$30K) with some exceeding \$100K.
\textbf{Lengler} and \textbf{portfolio} position themselves mainly around low payload (approx. 150 kg) and cluster together, shown in a dotted circle.

\figurename~\ref{fig:model_frontier_B2}--(b) plots the total cost vs. area coverage, where the farm size appears as a red dotted line, and the budget line appears as a blue dotted line.
The \texttt{PA} sits close to the farm size line, sufficient area coverage (\figurename~\ref{fig:model_frontier_B2}--(b)-(ii)) with varied costs.
\textbf{simulated annealing} extends beyond the minimum farm size, reaching sufficient area coverage (\figurename~\ref{fig:model_frontier_B2}--(b)-(ii)) with costs upto approx. 95K.
\textbf{BO} and \textbf{genetic algorithm} show a similar trend where area coverage increases with cost.
\textbf{random search} achieves moderate area coverage of about 4100–8500 m\textsuperscript{2} and stays under budget.
\textbf{discrete search} shows only one configuration at approx. 10000 m\textsuperscript{2}.
\textbf{PG-DSE} explores a broader solution space using evolutionary operators, which can momentarily include configurations in insufficient area coverage (\figurename~\ref{fig:model_frontier_B2}--(b)-(i)).
\textbf{Lengler} and \textbf{portfolio} cluster near the farm size line and have similar cost, showing a similar trend between these optimizers.

\begin{table}
    \scriptsize
	\centering
	\begin{tabular}{|p{3.9cm}|>{\centering\arraybackslash}p{1.8cm}|>{\centering\arraybackslash}p{1.7cm}|}
		\hline
		\textbf{State-of-the-Art Optimizers and Proposed Optimizer} & \textbf{Case Study 1: \$100K, 1 Acre} & \textbf{Case Study 2: \$1M, 10 Acres} \\
		\hline
		Simulated Annealing (DESTION 23) & 0.325 & 0.465 \\
		Bayesian Optimization (DESTION 23) & 0.498 & N/A  \\
        Random Search (DESTION 22) & 0.496 & 0.475 \\
		Genetic Algorithm (DESTION 22) & 0.503 & 0.527 \\
        Discrete Search (DESTION 22) & 0.307 & 0.600 \\
        Lengler (DESTION 22) & 0.634 & 0.686 \\
        Portfolio (DESTION 22) & 0.634 & 0.686 \\
        PG-DSE (ASP-DAC 23) & 0.619 & 0.360 \\
        Proposed Approach & 0.634 & 0.686 \\
		\hline
	\end{tabular}
    \vspace{-5pt}
	\caption{Weighted-score comparison for optimizer evaluation in Case Study 1 and Case Study 2.}
    \vspace{-20pt}
	\label{table:model_W}
\end{table}

\tablename~\ref{table:model_W} presents the weighted-score comparison for optimizer model evaluation, using the 
ROC weighting scheme described in ~\ref{Weighted Score Design}. 
In Case Study 1, the \texttt{PA}, \textbf{Lengler}, and \textbf{portfolio}  top the ranking by striking the best balance among cost, area coverage, and payload. 
\textbf{PG-DSE} ranks next highest, driven by multiple low-cost configurations, though some solutions incur budget penalties or provide limited area coverage.
\textbf{genetic algorithm}
claims the next highest scores, leveraging strong area coverage with controlled spending. 
\textbf{BO}, \textbf{Random search} and \textbf{simulated annealing} follow, as they identify feasible configurations but exhibit a higher cost for similar area coverage or payload compared to \textbf{genetic algorithm}.
\textbf{Discrete search} earns a slightly lower score due to limited designs.

 In Case Study 2, the \texttt{PA}, \textbf{Lengler}, and \textbf{portfolio} achieve the highest overall score by effectively balancing payload, area coverage, and cost. 
\textbf{discrete search} follow at a distance, \textbf{genetic algorithm} with  close behind, followed by \textbf{random search} and \textbf{simulated annealing}.
\textbf{PG-DSE} scores the lowest due to budget penalties or limited area coverage.
Also, \textbf{BO} from DESTION 23~\cite{Yu2023} gives no results since we lack hyperparameters (e.g., optimization calls, acquisition settings) that were unavailable, forcing untuned, general-purpose runs that cannot uncover viable configurations under tight constraints.

The trends shown in \figurename~\ref{fig:model_frontier_B2}
align well with the weighted scores in \tablename~\ref{table:model_W}. 
\texttt{PA} consistently shows strong trade-offs across cost, payload, and area coverage, explaining its top score in both weighted scores and plots.
\textbf{genetic algorithm}, \textbf{BO} and \textbf{random search} follow a similar pattern by keeping costs balanced while reaching practical area coverage. 
\textbf{discrete search}, and \textbf{simulated annealing} show moderate performance with lower scores.
Although \textbf{PG-DSE} scores a high weighted score, it has configurations that exceed the budget or provide limited area coverage.
\textbf{Lengler} and \textbf{portfolio} perform similarly, with overlapping plots and identical scores.

\subsection{Verification of DSE Results}

\begin{wraptable}{R}{6.22cm}
\vspace{-10pt}
\scriptsize
\centering
\begin{tabular}{|p{2.08cm}|>{\centering\arraybackslash}p{0.793cm}|>{\centering\arraybackslash}p{0.793cm}|>{\centering\arraybackslash}p{0.793cm}|}
\hline
\textbf{Method Verified by SAT} & \textbf{Total Configs} & \textbf{Valid Configs} & \textbf{Invalid Configs} \\
\hline
PA        & 18 & 18 & 0  \\
AP        & 20 & 17  & 3 \\
CA        & 18 & 18 & 0  \\
PC        & 13 & 0  & 13 \\
Simulated Annealing   & 12  & 12  & 0  \\
BO                    & 20 & 20 & 0  \\
PG-DSE                & 20 & 4 & 16 \\
Random Search         & 20  & 20  & 0  \\
Genetic Algorithm     & 20 & 20 & 0  \\
Discrete              & 1 & 1 & 0  \\
Lengler               & 18 & 18 & 0  \\
Portfolio             & 4  & 4  & 0  \\
\hline
\end{tabular}
\caption{SAT-based constraint verification across optimization methods, for Case Study 1.}
\vspace{-18pt}
\label{table:verification_results}
\end{wraptable}
\tablename~\ref{table:verification_results} summarizes the results of constraint verification
by our SAT-based approach described in Section~\ref{SAT based Verification} for Case Study 1.
The following approaches return only valid configurations without any invalid ones:
\texttt{PA}, \texttt{CA}, \textbf{BO}, \textbf{discrete}, \textbf{Lengler}, \textbf{random search}, \textbf{genetic algorithm}, \textbf{portfolio}, and \textbf{simulated annealing}.
In contrast, \texttt{PC} completely fails, returning all invalid results (13 out of 13), while \texttt{AP} and \textbf{PG-DSE} return 3 and 16 invalid configurations, respectively, out of 20 configurations. This highlights the strength of our ILP-based optimizer and the importance of verification in eliminating infeasible designs. 

\subsection{Execution Time Analysis}

Our approach can run on commercial off-the-shelf personal computers, such as laptops.
We measure the execution time of our approach on a computer with an AMD Ryzen 7 processor (8 cores @ 3.6 GHz), 16GB RAM, Windows 11, and Python 3.10.
In Case Study 1, \texttt{PA} requires 2.5 s to execute.
\textbf{PG-DSE} completes in approximately 487 ms, while both \textbf{genetic algorithm} and \textbf{portfolio} finish in about 10 ms. 
Each of \textbf{Lengler}, \textbf{random search}, \textbf{simulated annealing}, and \textbf{discrete search} takes up to 5 ms to complete.
In contrast, \textbf{BO} takes the longest time at 35.3 s.
Also, our SAT solver completes verification almost instantly, in just 8.6 ms.




\begin{table}[b]
\centering
\scriptsize
\vspace{-10pt}
\begin{tabular}{|p{1.2cm}|>{\centering\arraybackslash}p{3.2cm}|>{\centering\arraybackslash}p{0.75cm}|>{\centering\arraybackslash}p{0.8cm}|>{\centering\arraybackslash}p{0.7cm}|}
\hline
\textbf{Platform} & \textbf{Hardware Configurations} & \textbf{Payload (kg)} & \textbf{Runtime (hours)} & \textbf{Cost (USD)} \\
\hline
Rover & Plastic large body, large motor, medium battery, RPi 4B & 34.67 & 123.23 & 3449.56 \\
\hline
Drone & metal body, large motor, large battery, RPi~4B & 0.1 & 0.22 & 2345.34 \\
\hline
\end{tabular}
\vspace{-5pt}
\caption{Prototype hardware specifications derived from Case Study~2 solver outputs, with the drone implemented using a carbon-fiber body instead of the solver-selected metal body.}
\label{table:prototype_specs}
\end{table}

\subsection{Hardware Prototype Implementation}
\begin{wrapfigure}{r}{5.1cm}
\centering	
\vspace{-10pt}
\includegraphics[width=0.50\columnwidth]{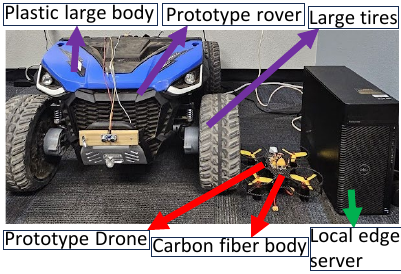}
\vspace{-5pt}
	\caption{Our prototype hardware constructed in line with the proposed approach's outputs.
    }
    \vspace{-16pt}
	\label{fig:Prototype_hardware}
\end{wrapfigure}

To demonstrate the feasibility of our DSE approach for precision agriculture, we build and verify a prototype UAV-UGV platform derived directly from the solver outputs of the proposed method, as shown in 
\figurename~\ref{fig:Prototype_hardware} and \tablename~\ref{table:prototype_specs}.

Our prototype consists of a rover, a drone, and a local edge server.
The rover uses a Raspberry Pi (RPi) 4B and a large plastic chassis with tires, providing sufficient runtime while supporting necessary sensors.
The drone (Duckiedrone DD24-B~\cite{Duckietown}) follows the solver-selected configuration, except for the body. While the solver specifies a metal body, we use a carbon-fiber frame for the lightweight, while keeping all other hardware components unchanged.
Also, a local edge server (\$2,000) provides off-board computation. 
Together, these elements confirm that ILP+SAT outputs are not abstract platforms but realizable, low-cost agricultural systems.


\section{Conclusion}
In summary, we propose a cost-aware DSE framework for multimodal UAV–UGV platforms in precision agriculture. 
By formulating the problem as an ILP with SAT verification, we address the challenges of balancing cost, coverage, and payload under strict resource constraints, while outperforming existing methods in terms of feasibility, cost, and the number of alternative configurations. 
Limitations, scope, and future work of this paper are further discussed below.

\noindent
\textbf{Limitations and Scope:}
To optimize the ILP, our framework targets static, deterministic conditions.
These choices streamline the ILP but limit its ability to reflect soil conditions, weather patterns, and long-term degradation. 
Communication and sensing are modeled at a high level, which can miss interference effects or routing complexity in large farms. 
Linear approximations of non-linear behaviors such as battery discharge, weight–efficiency decay, and torque–payload scaling further narrow realistic behavior. 

\noindent
\textbf{Future Work:}
We plan to extend our framework 
to other CPS domains with non-linear trade-offs, such as warehouse monitoring and disaster response. 
These settings introduce challenges like coverage vs. communication, mobility vs. energy, and safety vs. latency. 
We continue modeling these within an ILP and verifying feasibility with SAT.
Although each domain requires its own constraint set and weighting, the same systematic exploration process remains, 
making this work a foundation for broader CPS optimization.
\section*{Acknowledgment}
This work was supported in part by the National Science Foundation (NSF) under grants \#2449200 (NSF I/UCRC for IDEAS) and POSE-\#2449200 (An Open-Source Ecosystem to Coordinate Integration of Cyber-Physical Systems).


\bibliographystyle{IEEEtran}
\bibliography{library}

\end{document}